\documentclass[11pt,a4paper]{article}
\usepackage{jheppub}

\usepackage{amsmath,amssymb}
\usepackage{graphicx}
\usepackage{breakurl}

\graphicspath{ {figures/} }

\newcommand{\POWHEG}{{\tt POWHEG}}
\newcommand{\POWHEGBOX}{{\tt POWHEG\;BOX}}
\newcommand{\PYTHIA}{{\tt PYTHIA}}

\makeatletter
\renewcommand\@fpheader{\hfill \parbox{3cm}{MZ-TH/12-38}}
\renewcommand\@journal{}
\makeatother

\title{Slepton pair production in the \POWHEGBOX{}}

\author{Barbara~J\"ager,}
\author{Andreas~von~Manteuffel,}
\author{Stephan~Thier}

\affiliation{
  PRISMA Cluster of Excellence, 
  Institute of Physics (THEP),
  Johannes Gutenberg University, 
  55099 Mainz, Germany}

\emailAdd{jaegerba@uni-mainz.de}
\emailAdd{manteuffel@uni-mainz.de}
\emailAdd{thiers@uni-mainz.de}

\abstract{
We present an implementation for slepton pair production at hadron
\mbox{colliders} in the \POWHEGBOX{}, a framework for combining
next-to-leading order QCD calculations with parton-shower Monte-Carlo
programs.
Our code provides a SUSY Les Houches Accord interface for setting
the supersymmetric input parameters.
Decays of the sleptons and parton-shower effects are simulated
with {\tt PYTHIA}.
Focussing on a representative point in the supersymmetric parameter space
we show results for kinematic distributions that can be observed experimentally.
While next-to-leading order QCD corrections are sizable for all distributions,
the parton shower affects the color-neutral particles only marginally.
Pronounced parton-shower effects are found for jet distributions.
}

\notoc

\begin{document}

\maketitle

\newpage

%
\section{Introduction}
With the start-up of the CERN Large Hadron Collider (LHC) unprecedented
opportunities have emerged for experimentally accessing the terascale.
The capability of the LHC was impressively demonstrated by the recently
announced discovery~\cite{atlas-higgs,cms-higgs} of a new particle compatible
with the Higgs boson of the Standard Model.
Yet, numerous extensions of the Standard Model are being discussed, some
of the best-motivated ones being supersymmetric (SUSY) models that predict
the existence of supersymmetric partners for all Standard Model particles.

Direct production of the scalar partners of the leptons proceeds via
slepton pair production, if R-parity is conserved.
Compared to the production of the color-charged squarks and gluinos,
the production cross sections are smaller but also the signatures in
the detector are cleaner (see, e.g.,  Ref.~\cite{Kramer:2012bx} for a
recent review).
Current searches at ATLAS~\cite{atlas-sleptons} already supplement the
LEP limits~\cite{lep-sleptons} by excluding masses of the left-handed
sleptons up to 185~GeV for exclusive decays into a lepton and the lightest
neutralino, depending on the neutralino mass.
Precision measurements in the slepton sector would be
possible at a future $e^+e^-$ linear collider~\cite{Martyn:1999tc,Freitas:2003yp}
and to a less general extent also at a muon collider~\cite{Freitas:2011ti}.

For slepton pair production at hadron colliders,
the leading order (LO)~\cite{delAguila:1990yw,Baer:1993ew} and
next-to-leading order (NLO) QCD~\cite{Baer:1997nh} and
SUSY-QCD~\cite{Beenakker:1999xh}  contributions are known for the total
cross sections as well as for differential distributions.
The latter have been implemented in the computer package
{\tt PROSPINO}~\cite{Beenakker:1996ed}, which is publicly available.
Uncertainties inherent to the fixed-order parton-level calculations have
been further reduced by taking resummation effects into account.
In Refs.~\cite{Bozzi:2006fw}~and~\cite{Bozzi:2007qr}, transverse momentum
and threshold resummation for slepton pair production at hadron colliders
was considered, while Ref.~\cite{Bozzi:2007tea} provided joint resummation
predictions for this class of reactions.
More recently, using methods of soft-collinear effective theory 
invariant-mass distributions and total cross sections have been presented
at next-to-next-to-next-to-leading logarithmic accuracy~\cite{Broggio:2011bd}.

Ideally, precise predictions for the production cross sections should
be combined with realistic simulations of parton-shower effects,
multi-parton interactions, and underlying event -- effects that are
omni-present at hadron colliders.
Moreover, the complex decay patterns of the heavy SUSY particles should
fully be taken into account and realistic analysis cuts should be
applied.
These requirements call for the use of multi-purpose Monte-Carlo
generators such as {\tt HERWIG}~\cite{Marchesini:1991ch,Corcella:2000bw}
or {\tt PYTHIA}~\cite{Sjostrand:2006za}, which are, however, mostly
restricted to leading order matrix elements for hard scattering
processes.
A framework that allows to combine the merits of both, flexible
parton-shower programs with precise NLO-QCD calculations, is the
so-called \POWHEG{} approach~\cite{Nason:2004rx,Frixione:2007vw}.
The \POWHEGBOX{}~\cite {Alioli:2010xd} is a public tool providing all
general building blocks of this method, but requiring the user to
individually implement  process-specific pieces such as matrix elements
for the hard scattering process at NLO-QCD accuracy. 
Several Standard-Model processes bearing similar kinematic features as
slepton pair production are available in the
\POWHEGBOX{}~\cite{Alioli:2008gx,Frixione:2007nw}.
Selected SUSY processes have been studied
in the {\tt POWHEG} approach as well~\cite{Bagnaschi:2011tu,Klasen:2012wq}. 

In this article, we present an implementation of charged slepton pair
production at hadron colliders in the \POWHEGBOX{} and study the
relevance of parton-shower effects for representative distributions.
The slepton decay implementation of \PYTHIA{} is used to produce 
experimentally accessible final states.
Firstly, we describe our calculation in Sec.~\ref{sec:calc}.
In Sec.~\ref{sec:num}, we provide representative phenomenological
results for a specific parameter point of the minimal supersymmetric
extension of the Standard Model (MSSM).
Our conclusions are given in Sec.~\ref{sec:conc}.


\section{Framework of the calculation}
\label{sec:calc}
The NLO-QCD and SUSY-QCD (SQCD) corrections to slepton pair production at 
hadron colliders are known for quite some time 
\cite{Baer:1997nh,Beenakker:1999xh} and publicly available, e.~g.\ in
 the parton-level Monte Carlo program {\tt PROSPINO}~\cite{Beenakker:1996ed}.
Nonetheless, we decided to re-calculate the full $\mathcal{O}(\alpha_s)$
 corrections to this class of processes in the framework of the MSSM.
In the following, we consider the production of a pair of charged sleptons
and leave a discussion of final states involving sneutrinos for future work.

\begin{figure}
\centerline{
\includegraphics[width=1.0\textwidth,clip]{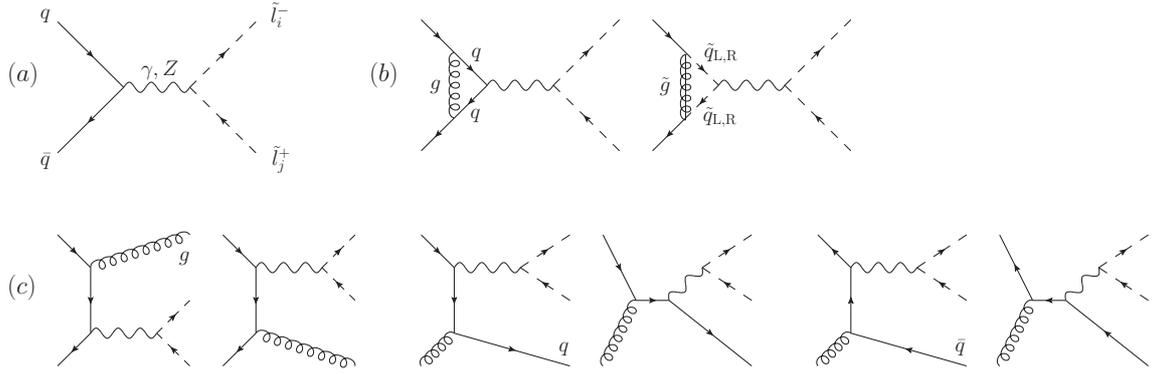}
}
\caption{\label{fig:feynman}
Feynman diagrams contributing to slepton pair production at
hadron colliders: tree-level contribution (a),
virtual NLO-QCD and SQCD corrections (b) and
real-emission contributions (c) from different channels.
}
\end{figure}

At LO, the production of charged slepton pairs at hadron colliders
dominantly proceeds via the Drell-Yan type annihilation of a quark-antiquark
pair into a virtual $Z$~boson or photon that in turn decays into a pair 
of heavy sfermions, c.f.\ Fig.~\ref{fig:feynman}~(a).
The SQCD-NLO corrections to this reaction comprise virtual one-loop corrections
and real-emission contributions with an extra parton in the final state.
To the former only self-energy and vertex corrections contribute, since the
sleptons do not carry color charge, see Fig.~\ref{fig:feynman}~(b).
For the real-emission corrections we consider $q\bar q$ annihilation
diagrams with a gluon being radiated off either of the incoming fermions,
and crossing-related contributions where the gluon is promoted to the
initial state while the final  state features a quark or anti-quark in
addition to the slepton pair, as shown in Fig.~\ref{fig:feynman}~(c).

We employ the on-shell scheme for the renormalization of the wave functions 
of the massless quarks, which requires the evaluation of additional self 
energy diagrams not depicted here.
Other quantities such as the couplings receive no non-trivial contributions 
at the order of perturbation theory considered.
We account for large universal corrections to the electromagnetic
coupling due to light fermions by a running coupling.
We input its value at the scale $m_Z$
and use the resummed one-loop contributions to 
evolve to the invariant mass of the slepton pair.

We generate the Feynman diagrams with {\tt QGRAF}~\cite{Nogueira:1991ex} 
and build the interference terms with in-house developed 
{\tt FORM}~\cite{Vermaseren:2000nd,Kuipers:2012rf} scripts. 
The relative sign of interference produced by {\tt QGRAF} cannot be used 
due to the presence of Majorana fermions; we generate this sign according 
to Ref.~\cite{Denner:1992vza}.
Contraction of Lorentz indices leaves us with scalar integrals which are 
reduced to master integrals with the program 
{\tt Reduze\,2}~\cite{Studerus:2009ye,vonManteuffel:2012yz}.
The resulting expressions are exported to {\tt Fortran\,77} files that make
use of the  {\tt QCDLoop} library~\cite{Ellis:2007qk} for the numerical
evaluation of the scalar master integrals.
Conventional dimensional regularization is used to regularize both
ultraviolet (UV) and infrared (IR) divergences. No supersymmetry-restoring 
counterterms are needed here~\cite{Hollik:2001cz}.
The calculation employs $\gamma^5 = (i/4!) \epsilon_{\mu\nu\rho\sigma}
\gamma^\mu\gamma^\nu\gamma^\rho\gamma^\sigma$ for the axial
couplings~\cite{'tHooft:1972fi,Larin:1993tq}.
Additionally we performed an independent calculation utilizing the 
{\tt Mathematica} package {\tt FeynArts}~\cite{Hahn:2000kx} for 
amplitude generation, {\tt FormCalc}~\cite{Hahn:1998yk,Hahn:2006zy} for 
algebraic simplifications including tensor reductions, and 
{\tt LoopTools}~\cite{Hahn:1998yk,vanOldenborgh:1990yc} for the 
numerical evaluation of the scalar integrals.
This calculation uses the so-called naive 
anticommuting scheme~\cite{Chanowitz:1979zu} for $\gamma^5$.
The results of these two calculations are in complete agreement with each
other.
In both approaches, we neglect masses of quarks and leptons for the first 
and second generations.
Consequently, the scalar partners of these left and right-chiral fermions 
are treated as mass eigenstates.
While partons of the third generation may safely be ignored in the initial 
state, mixing of staus is taken into account.

Having calculated the $\mathcal{O}(\alpha_s)$ corrections to the partonic 
scattering process has put us in a position to work out an implementation 
of slepton pair production in the \POWHEGBOX{}. We merely have to provide 
the following ingredients:
\begin{itemize}
\item
a list of all flavor structures for the LO and the real-emission contributions, 
\item
a suitable parameterization of the Born phase space, 
\item 
the Born amplitudes squared for all partonic subprocesses,
\item
the color correlated and the spin correlated Born amplitudes,
\item 
the finite part of the virtual QCD and SQCD corrections,
\item 
the real-emission matrix elements squared for all partonic subprocesses.
\end{itemize}
Once these building blocks are implemented in the \POWHEGBOX{}, the program 
itself performs the phase-space integration and convolution with the set of 
parton-distribution functions selected by the user.
IR divergent configurations are taken care of internally by a subtraction 
procedure based on Ref.~\cite{Frixione:1995ms}. The \POWHEGBOX{}  also 
provides the means for checking that the real-emission contributions approach 
the respective counterterms for soft and collinear configurations. 
This constitutes a useful debugging tool  as it tests the correct
normalization of the real-emission amplitudes
relative to the Born matrix elements that are used for the computation 
of the counterterms.

We have compared our total cross section with results
obtained from {\tt PROSPINO 2.1} and found agreement both
at LO and at NLO.
In addition, we checked that our virtual Standard-Model-QCD and SUSY-QCD corrections
as well as LO and NLO total cross sections are in agreement with \cite{Broggio:2011bd}. 

In our implementation, spectrum dependent decays of the on-shell sleptons
are incorporated via {\tt PYTHIA 6.4.25}. 
For the input of fully general MSSM parameters, we employ the 
SUSY Les Houches Accord ({\tt SLHA})~\cite{Skands:2003cj,Allanach:2008qq}
interface.
The production part of our implementation utilizes the {\tt SLHALib\,2} 
library~\cite{Hahn:2006nq} ({\tt SLHA\,1 \& SLHA\,2}), while the decay code
of \PYTHIA{}  already features configuration via {\tt SLHA\,1}.
In this way, the output of spectrum 
generators can be passed directly to our program in a well-defined and 
user-friendly way. 


\section{Phenomenological results and discussion}
\label{sec:num}
Our implementation of slepton pair production in the \POWHEGBOX{} will be made
publicly available at the homepage of the \POWHEGBOX{} project,
\url{http://powhegbox.mib.infn.it}.
We encourage the phenomenologist to use
this code with his or her own
preferred settings for the SUSY spectrum, Standard-Model input parameters
and experimental selection cuts.
A documentation of the code including recommended values for technical
parameters is provided together with the code.
Here, we show representative results for a specific point in the SUSY
parameter space. 

We consider proton-proton collisions at the LHC with a center-of-mass
energy of $\sqrt{S}=8$~TeV.
For the electroweak parameters we use input
provided by the particle data group (PDG)~\cite{PDG2012}.
We set the mass of the $Z$~boson $m_Z = 91.1876$~GeV, the fine-structure constant
$\alpha = 1/137.036$ and use a running electromagnetic coupling in the
on-shell scheme with $\alpha(m_Z) = 1/128.919$.
For the electroweak mixing angle we obtain $\sin^2\theta_W=0.23103$ from
the PDG value for Fermi's constant, $m_Z$ and $\alpha(m_Z)$ using the
tree-level relation.
The $Z$~width is set to its PDG value but has no impact on the results discussed
in the following.
For the masses of the sleptons and the lightest neutralino we set
\begin{align}
m_{\tilde{l}_R} &= 180~\text{GeV}\qquad
 \text{for~}\tilde{l}_R = \tilde{e}_R, \tilde{\mu}_R,\nonumber\\
m_{\tilde{\chi}^0_1}&=80~\text{GeV},
\end{align}
while for squarks and gluino we use
 $m_{\tilde{q}_L}=m_{\tilde{q}_R}=1500$~GeV
 for $\tilde{q}=\tilde{u},\tilde{d},\tilde{c},\tilde{s}$ and
 $m_{\tilde{g}}=2000$~GeV.
While SUSY particles of these masses are not excluded by collider data,
from Ref.~\cite{atlas-sleptons} one may expect such a slepton sector
to be accessible by future slepton searches at the LHC experiments.
The virtual SQCD corrections induced by the squarks and gluinos are
included but found to be numerically irrelevant because of their large
masses in this setup.

For the numerical analysis we restrict ourselves to the pair production of
the SUSY partners of the right-handed leptons of the first and second
generation, the R-selectrons and R-smuons.
The sleptons are assumed to exclusively decay into a lepton and the lightest
neutralino, where we employ {\tt PYTHIA 6.4.25} for the decay. Throughout, we
switch off QED radiation, underlying event and hadronization effects
in \PYTHIA.
Unless stated otherwise, factorization and renormalization scales are
identified with the invariant mass of the slepton pair.
For the parton-distribution functions of the proton, we use the MSTW2008
parameterization~\cite{Martin:2009iq} as implemented in the
{\tt LHADPDF}~library~\cite{Whalley:2005nh}.
The real-emission contributions of the NLO-QCD calculation as well as the
parton shower can give rise to partons in the final state.
These are recombined into jets according to the anti-$k_T$
algorithm~\cite{Cacciari:2008gp} as implemented in the {\tt FASTJET}
package~\cite{Cacciari:2005hq,Cacciari:2011ma}, with $R=0.4$ and
$y^\mathrm{jet}<4.5$. 

For the total cross section we find $\sigma = 5.93$~fb at NLO-QCD.
In order to illustrate the dependence of our results on unphysical scales,
we evaluated the total cross sections for our default
setup at LO and NLO-QCD, varying the 
factorization and renormalization scales in the range $\mu_0/2$ to $2 \mu_0$, 
with $\mu_0 = 2 m_{\tilde{l}_R}$.
Note that the renormalization scale does not enter the LO results, since at
tree-level slepton pair production is a purely electroweak process.
While at LO we find a scale uncertainty of about 8\%, the NLO results for
different values of $\mu_0$ differ by only 4\%.

\begin{figure}[t]
\centerline{
\includegraphics[width=0.8\textwidth,clip]{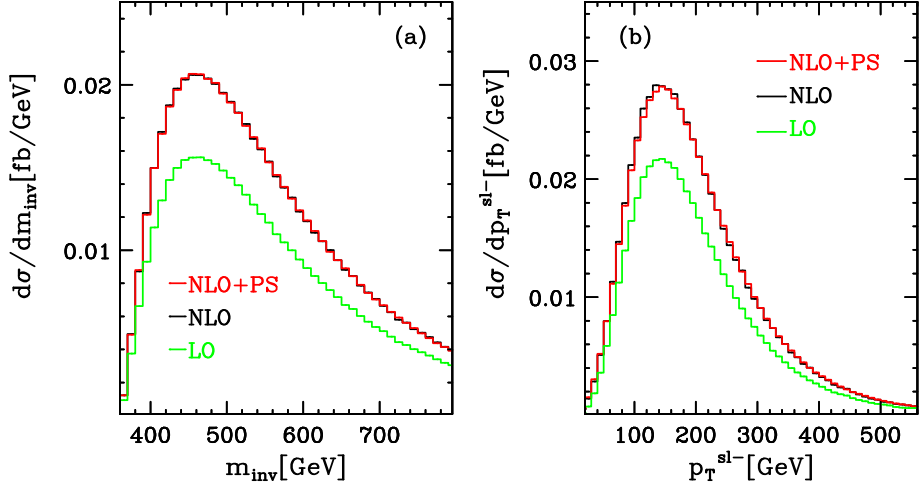}
}
\caption{\label{fig:mss}
Invariant-mass distribution of the slepton pair~(a) and
transverse momentum of the negatively charged selectron~(b) at LO (green),
NLO-QCD (black), and with {\tt POWHEG+PYTHIA} (red) for our default 
LHC setup with $\sqrt{S}=8$~TeV.
}
\end{figure}

In Figs.~\ref{fig:mss}~(a) and (b) we present the invariant-mass distribution of the
slepton pair and the transverse momentum of the negatively charged selectron,
respectively, for our default setup at LO, NLO, and for {\tt POWHEG+PYTHIA}
(NLO+PS).
The statistical errors on the distributions are at the order of 1\% per
bin or smaller and therefore not explicitly indicated in the histograms. 
As expected,  relatively large positive NLO-QCD corrections are encountered
for both observables, while the impact of the parton shower on the NLO result
is marginal for these slepton distributions.

\begin{figure}[t]
\centerline{
\includegraphics[width=0.8\textwidth,clip]{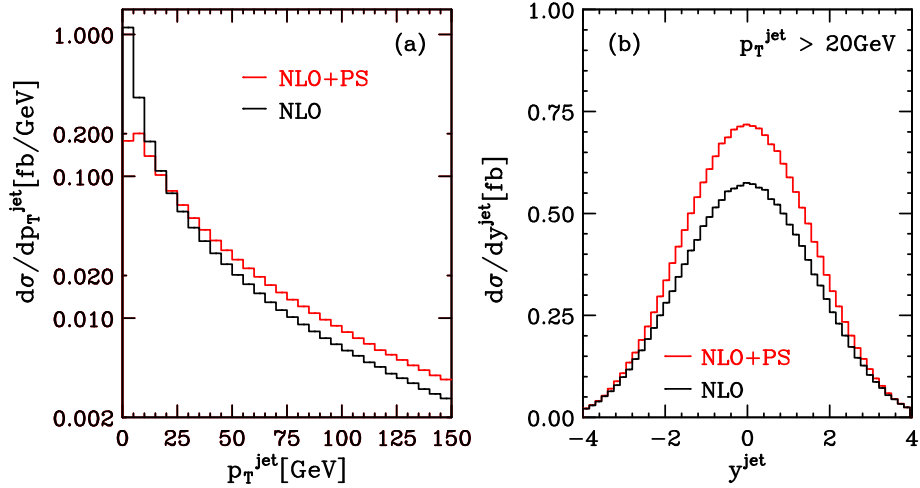}
}
\caption{\label{fig:jet}
Transverse momentum (a) and rapidity distribution (b) of the hardest jet
for slepton pair production at the LHC with $\sqrt{S}=8$~TeV.
In (b) we additionally require $p_T^{jet}>20$~GeV.  
}
\end{figure}

More pronounced parton-shower effects can be observed in distributions related
to the hardest jet produced in association with the slepton pair.
In the NLO-QCD calculation, a jet can only arise from the
final-state parton of the real-emission contributions.
In contrast,  the parton shower can give rise to events
with even more than one hard jet in the {\tt POWHEG+PYTHIA} result.
The transverse-momentum distribution of the hardest jet is shown in
Fig.~\ref{fig:jet}~(a) for the fixed-order and the  {\tt POWHEG+PYTHIA} calculation. 
While in the fixed-order QCD prediction $d\sigma/dp_T^\mathrm{jet}$
rises steadily towards very low transverse momenta, this rise is damped
by the Sudakov factor in the {\tt POWHEG+PYTHIA} result. 
In either case, the extra jet tends to be produced at central rapidities,
as illustrated by Fig.~\ref{fig:jet}~(b).
There, we imposed an extra cut on the transverse momentum of the jet,
$p_T^\mathrm{jet}>20$~GeV to avoid contributions from very soft partons
produced by the parton shower.

\begin{figure}[t]
\centerline{
\includegraphics[width=0.8\textwidth,clip]{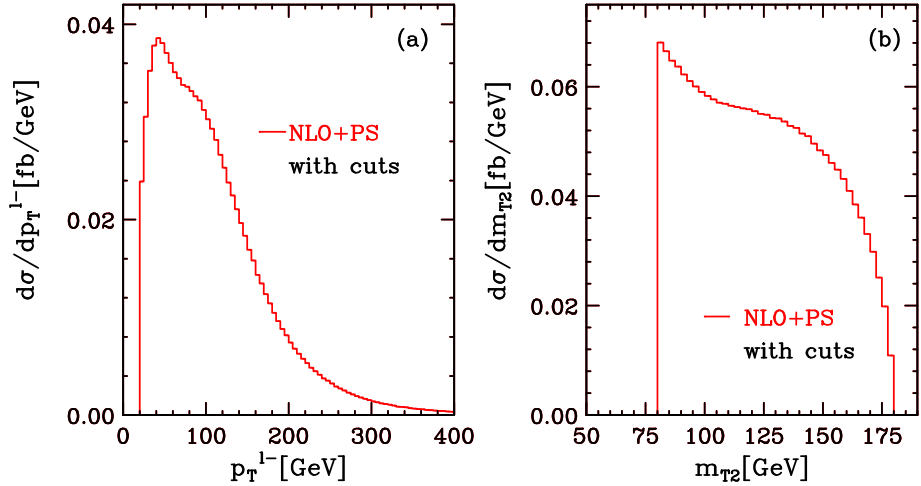}
}
\caption{\label{fig:lep}
Transverse-momentum distribution of the negatively charged lepton~(a) and $m_{T2}$~distribution~(b) 
with {\tt POWHEG+PYTHIA} within the cuts of Eq.~(\ref{eq:lepton-cuts}) for slepton pair
production at the LHC with $\sqrt{S}=8$~TeV. 
}
\end{figure}

An important feature of our code is the possibility to interface the
calculation of the slepton pair production cross section with \PYTHIA{}
in such a way that decays of the sleptons can be simulated, and the user
gets access on kinematic distributions of the decay products within
selection cuts of his own choice.
In Fig.~\ref{fig:lep}~(a)
we present predictions at the NLO+PS level for the transverse-momentum
distribution of the hardest negatively charged lepton within our standard
setup, and with additional cuts on the charged leptons,    
requiring them to be hard and centrally produced, 
\begin{equation}
\label{eq:lepton-cuts}
p_T^\ell > 20~\mathrm{GeV}\quad \text{and}\quad 
\left|\eta^\ell\right| < 2.5\,. 
\end{equation}
Providing predictions in terms of final decay products,
our code can be used directly also for more elaborate analysis techniques
such as studies of the $m_{T2}$ distribution defined in
Ref.~\cite{Lester:1999tx}.
This is demonstrated in Fig.~\ref{fig:lep}~(b), where we used the code
of Ref.~\cite{Cheng:2008hk} for the calculation of $m_{T2}$.

While these results are meant as an illustration of the capability of our
code, we would like to stress that the user is free to add distributions,
modify cuts, and set masses of the SUSY particles according to his needs
in a straightforward manner. 


\section{Conclusions}
\label{sec:conc}
In this work, we have presented an implementation of slepton pair production 
at hadron colliders in the \POWHEGBOX{}, a framework that allows to combine 
NLO-QCD calculations with parton-shower Monte-Carlo programs.
We have described how we calculated the QCD and  SQCD corrections to the 
hard scattering process, and then explained the technical details of their 
implementation in a publicly available computer code. 
The program we developed contains an {\tt SLHA} interface
that allows the use of MSSM input parameters provided by
stand-alone spectrum generators.
When our program is run with \PYTHIA{}, decays of the sleptons can be
simulated by the Monte-Carlo generator. 
 
We provided representative results for various kinematic distributions
of the sleptons and their decay products for a specific point in the
MSSM parameter space. 
Our analysis showed that NLO-QCD corrections modify tree-level predictions
for slepton pair production processes significantly.
Parton-shower effects can be pronounced for jet observables, while they are
typically small for distributions related to the sleptons. 


\section*{Note added in proof}
During the completion of this work, a similar study
appeared~\cite{FridmanRojas:2012yh}, containing a simulation of slepton
pair production at NLO-QCD in {\tt HERWIG++}~\cite{Bahr:2008pv}.
Two separate implementations of this reaction in completely
different program packages provide excellent means to check the
independence of predictions from a specific tool.
In this sense, it would be interesting to compare the results of the
two implementations in the future. 

%
\acknowledgments{
We are grateful to Alessandro Broggio for numerous helpful discussions and to
\mbox{Christian} Speckner and Giulia Zanderighi for valuable comments. 
This work is supported in part by the Research Center {\em
  Elementary Forces and Mathematical Foundations (EMG)} of the
Johannes-Gutenberg-Universit\"at Mainz and by the
German Research Foundation (DFG).
S.~T.\ is a recipient of a fellowship through the graduate school
{\em Symmetry Breaking} (DFG/ GRK 1581). 
}



\begin{thebibliography}{99}

\bibitem{atlas-higgs}
  G.~Aad {\it et al.}  [The ATLAS Collaboration],
  arXiv:1207.7214 [hep-ex].

\bibitem{cms-higgs}
  S.~Chatrchyan {\it et al.}  [The CMS Collaboration],
  arXiv:1207.7235 [hep-ex].

\bibitem{Kramer:2012bx}
  M.~Kramer {\it et al.},
  arXiv:1206.2892 [hep-ph].

\bibitem{atlas-sleptons}
ATLAS collaboration, 
ATLAS-CONF-2012-076 (2012); http://cdsweb.cern.ch/record/1460273.

\bibitem{lep-sleptons}
LEP SUSY Working Group (ALEPH, DELPHI, L3, OPAL), Notes LEPSUSYWG/01-03.1 and
  04-01; http://lepsusy.web.cern.ch/lepsusy/Welcome.html.

\bibitem{Martyn:1999tc}
  H.~-U.~Martyn and G.~A.~Blair,
  in 2nd ECFA/DESY study (1998-2001) 743-747
  [hep-ph/9910416].

\bibitem{Freitas:2003yp}
  A.~Freitas, A.~von Manteuffel and P.~M.~Zerwas,
  Eur.\ Phys.\ J.\ C {\bf 34} (2004) 487
  [hep-ph/0310182].

\bibitem{delAguila:1990yw}
  F.~del Aguila and L.~Ametller,
  Phys.\ Lett.\ B {\bf 261} (1991) 326.

\bibitem{Freitas:2011ti}
  A.~Freitas,
  arXiv:1107.3853 [hep-ph].

\bibitem{Baer:1993ew}
  H.~Baer, C.~H.~Chen, F.~Paige and X.~Tata,
  Phys.\ Rev.\ D {\bf 49} (1994) 3283
  [hep-ph/9311248].

\bibitem{Baer:1997nh}
  H.~Baer, B.~W.~Harris and M.~H.~Reno,
  Phys.\ Rev.\ D {\bf 57} (1998) 5871
  [hep-ph/9712315].

\bibitem{Beenakker:1999xh}
  W.~Beenakker, M.~Klasen, M.~Kramer, T.~Plehn, M.~Spira and P.~M.~Zerwas,
  Phys.\ Rev.\ Lett.\  {\bf 83} (1999) 3780
   [Erratum-ibid.\  {\bf 100} (2008) 029901]
  [hep-ph/9906298].

\bibitem{Beenakker:1996ed}
  W.~Beenakker, R.~Hopker and M.~Spira,
  hep-ph/9611232.

\bibitem{Bozzi:2006fw}
  G.~Bozzi, B.~Fuks and M.~Klasen,
  Phys.\ Rev.\ D {\bf 74} (2006) 015001
  [hep-ph/0603074].
  
\bibitem{Bozzi:2007qr}
  G.~Bozzi, B.~Fuks and M.~Klasen,
  Nucl.\ Phys.\ B {\bf 777} (2007) 157
  [hep-ph/0701202].

\bibitem{Bozzi:2007tea}
  G.~Bozzi, B.~Fuks and M.~Klasen,
  Nucl.\ Phys.\ B {\bf 794} (2008) 46
  [arXiv:0709.3057 [hep-ph]].

\bibitem{Broggio:2011bd}
  A.~Broggio, M.~Neubert and L.~Vernazza,
  JHEP {\bf 1205} (2012) 151
  [arXiv:1111.6624 [hep-ph]].
     
\bibitem{Marchesini:1991ch}
G.~Marchesini {\it et al.}, 
Comp.\ Phys.\ Commun.\ {\bf 67} (1992) 465.
%
\bibitem{Corcella:2000bw}
  G.~Corcella {\it et al.},
  JHEP {\bf 0101 } (2001)  010.
  [hep-ph/0011363].

\bibitem{Sjostrand:2006za}
  T.~Sjostrand, S.~Mrenna, P.~Z.~Skands,
  JHEP {\bf 0605 } (2006)  026.
  [hep-ph/0603175]. 

\bibitem{Nason:2004rx}
  P.~Nason,
  JHEP {\bf 0411 } (2004)  040.
  [hep-ph/0409146].

\bibitem{Frixione:2007vw}
  S.~Frixione, P.~Nason, C.~Oleari,
  JHEP {\bf 0711 } (2007)  070.
  [arXiv:0709.2092 [hep-ph]].

\bibitem{Alioli:2010xd}
  S.~Alioli, P.~Nason, C.~Oleari, E. Re,
  JHEP {\bf 1006 } (2010)  043.
  [arXiv:1002.2581 [hep-ph]].

\bibitem{Alioli:2008gx}
  S.~Alioli, P.~Nason, C.~Oleari and E.~Re,
  JHEP {\bf 0807} (2008) 060
  [arXiv:0805.4802 [hep-ph]].

\bibitem{Frixione:2007nw}
  S.~Frixione, P.~Nason and G.~Ridolfi,
JHEP {\bf 0709} (2007) 126
[arXiv:0707.3088 [hep-ph]].

\bibitem{Bagnaschi:2011tu}
  E.~Bagnaschi, G.~Degrassi, P.~Slavich and A.~Vicini,
  JHEP {\bf 1202} (2012) 088
  [arXiv:1111.2854 [hep-ph]].

\bibitem{Klasen:2012wq}
  M.~Klasen, K.~Kovarik, P.~Nason and C.~Weydert,
  arXiv:1203.1341 [hep-ph].

\bibitem{Nogueira:1991ex}
  P.~Nogueira,
  J.\ Comput.\ Phys.\  {\bf 105} (1993) 279.


\bibitem{Vermaseren:2000nd}
  J.~A.~M.~Vermaseren,
  math-ph/0010025.

\bibitem{Kuipers:2012rf}
  J.~Kuipers, T.~Ueda, J.~A.~M.~Vermaseren and J.~Vollinga,
  arXiv:1203.6543 [cs.SC].

\bibitem{Denner:1992vza}
  A.~Denner, H.~Eck, O.~Hahn and J.~Kublbeck,
  Nucl.\ Phys.\ B {\bf 387} (1992) 467.

\bibitem{Studerus:2009ye}
  C.~Studerus,
  Comput.\ Phys.\ Commun.\  {\bf 181} (2010) 1293
  [arXiv:0912.2546 [physics.comp-ph]].

\bibitem{vonManteuffel:2012yz}
  A.~von Manteuffel and C.~Studerus,
  arXiv:1201.4330 [hep-ph],

\bibitem{Ellis:2007qk}
  R.~K.~Ellis and G.~Zanderighi,
  JHEP {\bf 0802} (2008) 002
  [arXiv:0712.1851 [hep-ph]].
  
\bibitem{Hollik:2001cz}
  W.~Hollik and D.~Stockinger,
  Eur.\ Phys.\ J.\ C {\bf 20} (2001) 105
  [hep-ph/0103009].

\bibitem{'tHooft:1972fi}
  G.~'t Hooft and M.~J.~G.~Veltman,
  Nucl.\ Phys.\ B {\bf 44} (1972) 189.


\bibitem{Larin:1993tq} 
  S.~A.~Larin,
  Phys.\ Lett.\ B {\bf 303}, 113 (1993)
  [hep-ph/9302240].

\bibitem{Hahn:2000kx}
  T.~Hahn,
  Comput.\ Phys.\ Commun.\  {\bf 140} (2001) 418
  [hep-ph/0012260].

\bibitem{Hahn:1998yk}
  T.~Hahn and M.~Perez-Victoria,
  Comput.\ Phys.\ Commun.\  {\bf 118} (1999) 153
  [hep-ph/9807565].

\bibitem{Hahn:2006zy}
  T.~Hahn,
  Comput.\ Phys.\ Commun.\  {\bf 178} (2008) 217
  [hep-ph/0611273].

\bibitem{vanOldenborgh:1990yc}
  G.~J.~van Oldenborgh,
  Comput.\ Phys.\ Commun.\  {\bf 66} (1991) 1.
  
\bibitem{Chanowitz:1979zu}
  M.~S.~Chanowitz, M.~Furman and I.~Hinchliffe,
  Nucl.\ Phys.\ B {\bf 159} (1979) 225.


\bibitem{Frixione:1995ms}
S.~Frixione, Z.~Kunszt, A.~Signer,
Nucl.\ Phys.\ {\bf B467} (1996) 399. 
[hep-ph/9512328]. 


\bibitem{Skands:2003cj}
  P.~Z.~Skands  {\it et al.},
  JHEP {\bf 0407} (2004) 036
  [hep-ph/0311123].

\bibitem{Allanach:2008qq}
  B.~C.~Allanach  {\it et al.},
  Comput.\ Phys.\ Commun.\  {\bf 180} (2009) 8
  [arXiv:0801.0045 [hep-ph]].

\bibitem{Hahn:2006nq}
  T.~Hahn,
  Comput.\ Phys.\ Commun.\  {\bf 180} (2009) 1681
  [hep-ph/0605049].
  
  
\bibitem{PDG2012}
 J.~Beringer et al.\ (Particle Data Group), Phys.\ Rev.\ D86 (2012) 010001.

\bibitem{Martin:2009iq}
  A.~D.~Martin, W.~J.~Stirling, R.~S.~Thorne, G.~Watt,
  Eur.\ Phys.\ J.\  {\bf C63 } (2009)  189-285.
  [arXiv:0901.0002 [hep-ph]].

\bibitem{Whalley:2005nh}
M.~R.~Whalley, D.~Bourilkov, R.~C.~Group, 
hep-ph/0508110. 

\bibitem{Cacciari:2008gp}
  M.~Cacciari, G.~P.~Salam and G.~Soyez,
  JHEP {\bf 0804} (2008) 063
  [arXiv:0802.1189 [hep-ph]].

\bibitem{Cacciari:2005hq}
M.~Cacciari, G.~P.~Salam, 
Phys.\ Lett.\ {\bf B641} (2006) 57
[hep-ph/0512210]. 

\bibitem{Cacciari:2011ma}
  M.~Cacciari, G.~P.~Salam and G.~Soyez,
  arXiv:1111.6097 [hep-ph].
  
\bibitem{Lester:1999tx}
  C.~G.~Lester and D.~J.~Summers,
  Phys.\ Lett.\ B {\bf 463} (1999) 99
  [hep-ph/9906349].

\bibitem{Cheng:2008hk}
  H.~-C.~Cheng and Z.~Han,
  JHEP {\bf 0812} (2008) 063
  [arXiv:0810.5178 [hep-ph]].

\bibitem{FridmanRojas:2012yh}
  I.~Fridman-Rojas and P.~Richardson,
  arXiv:1208.0279 [hep-ph].


\bibitem{Bahr:2008pv}
  M.~Bahr {\it et al.},
  Eur.\ Phys.\ J.\ C {\bf 58} (2008) 639
  [arXiv:0803.0883 [hep-ph]].

\end{thebibliography}
\end{document}